\documentclass[11pt]{article}
\usepackage{hyperref}
\begin{document}

\title{Fluid Flower :\\Microliquid Patterning via Drop Impact}

\author{Minhee Lee and Ho-Young Kim \\
\\\vspace{6pt} School of Mechanical and Aerospace Engineering, \\ Seoul National University, Seoul 151-744, Korea}

\maketitle

\begin{abstract}
In microfluidic technologies, direct patterning of liquid without
resorting to micromachined solid structures has various advantages
including reduction of the frictional dissipation and the
fabrication cost. This fluid dynamics video illustrates the method
to micropattern a liquid on a solid surface with drop impact. We
experimentally show that a water drop impacting with the
wettability-patterned solid retracts fast on the hydrophobic
regions while being arrested on the hydrophilic areas.

\end{abstract}

\section{Introduction}

The experimental videos are linked
\href{http://ecommons.library.cornell.edu/bitstream/1813/11451/2/Fluid%20Flower_Microliquid%20Patterning%20via%20Drop%20Impact%20mpeg2.m2v}{Video1}
and
\href{http://ecommons.library.cornell.edu/bitstream/1813/11451/1/Fluid%20Flower_Microliquid%20Patterning%20via%20Drop%20Impact%20mpeg1.m1v}{Video2}.
\\

In this video water is micropatterned on a glass surface which has
alternating patterns of hydrophilic and hydrophobic natures. The
surface pattern is made by photolithography and selective
deposition of decyltrichlorosilane self-assembled monolayers.
Water drops with the diameter of 3 mm impacting on the surface at
around 3 m/s exhibit aesthetically pleasing patterns while
recoiling due to capillary effects guided by the wettability
pattern. The drop motions are captured by a high-speed camera at
the rates between 2000 and 5130 frames per second.

\end{document}